\documentclass[showpacs,amsmath,amssymb,twocolumn,prl,superscriptaddress]{revtex4-1}
\usepackage{amssymb}
\usepackage[dvips]{graphicx}
\usepackage{enumerate}
\usepackage{epsfig}
\usepackage{subfigure}
\usepackage{xcolor}
\usepackage[T1]{fontenc}
\usepackage{fullpage}
\usepackage{amsthm,amsfonts,amssymb,amscd,mathrsfs,xspace,framed}
\usepackage{amsmath}
\usepackage{color}
\usepackage{setspace}
\usepackage{url}
\usepackage{wrapfig}
\usepackage{enumitem}
\usepackage{wrapfig}
\usepackage{verbatim}
\begin{document}

\title{Clarification of exceptional point contribution for photonic sensing}

\author{Dalton Anderson}
\affiliation{James C. Wyant College of Optical Sciences, The University of Arizona, Tucson, Arizona 85721, USA}
\author{Manav Shah}
\affiliation{James C. Wyant College of Optical Sciences, The University of Arizona, Tucson, Arizona 85721, USA}
\author{Linran Fan}
\email{lfan@optics.arizona.edu}
\affiliation{James C. Wyant College of Optical Sciences, The University of Arizona, Tucson, Arizona 85721, USA}

\begin{abstract}
Exceptional points, with simultaneous coalescence of eigen-values and eigen-vectors, can be realized with non-Hermitian photonic systems. With the enhanced response, exceptional points have been proposed to improve the performance of photonic sensing. Recently, there are intense debate about the actual sensing advantage of exceptional points. The major concern is that intrinsic noise is also amplified at exceptional points. Here, we aim to clarify the contribution of exceptional points for photonic sensing. This is achieved by analyzing the condition to realize divergent quantum Fisher information in linear non-Hermitian photonic systems. We show that the divergence of quantum Fisher information is the result of lasing threshold, instead of exceptional points. However, exceptional points correspond to the condition that lasing threshold is simultaneously achieved across multiple photonic modes. Therefore, exceptional points can further improve the sensitivity on top of lasing threshold. On the other hand, exceptional points alone cannot provide sensing advantage.
\end{abstract}

\maketitle
Optical metrology plays a critical role in modern precision measurement. Prominent examples include laser spectroscopy for molecule identification ~\cite{demtroder1982laser,solarz2017laser}, optical frequency comb for timing and ranging ~\cite{udem2002optical,diddams2010evolving,fortier201920}, and optomechanics for inertial navigation and force sensing~\cite{aspelmeyer2014cavity,metcalfe2014applications}. The capability of optical metrology is further strengthened by the recent demonstration of photonic exceptional points~\cite{el2018non,ozdemir2019parity,miri2019exceptional,wiersig2020review}. The initial rationale to use exceptional points for photonic sensing is that the frequency splitting $\delta\omega$ of coupled photonic cavities at $n$th-order exceptional points scales as $\delta\omega \propto\varepsilon^{1/n}$, in contrast to linear scaling $\delta\omega \propto\varepsilon$ without exceptional points ~\cite{ozdemir2019parity,miri2019exceptional,wiersig2020review}. Therefore, a stronger spectral response can be achieved at exceptional points.

With the enhanced signal response, it is widely accepted that the overall photonic sensing performance can be improved if the noise is dominated by later measurement stages~\cite{wiersig2020review}. As exceptional points are typically realized with non-Hermitian photonic systems, there are unavoidable intrinsic noise processes such as vacuum fluctuations due to loss and spontaneous emission in gain medium~\cite{gardiner2004quantum}. This has caused intense debate about whether exceptional points can still show sensing advantage if intrinsic noise processes are dominant~\cite{langbein2018no,lau2018fundamental,chen2019sensitivity,zhang2019quantum,wang2020petermann,smith2020beyond,wiersig2020robustness,kononchuk2022enhanced,duggan2022limitations}. The major concern is that intrinsic noise is also amplified due to the coalescence of eigen-vectors. This has been confirmed with theoretical~\cite{langbein2018no,lau2018fundamental}, analytical~\cite{duggan2022limitations}, and experimental~\cite{wang2020petermann} studies of specific measurement protocols of exceptional points. However, according to the quantum information theory, analysis built upon specific measurement protocols is not sufficient to draw a conclusion to the ultimate limit of photonic sensing systems~\cite{degen2017quantum}. Instead, quantum Fisher information (QFI), which is independent of specific measurement protocols, can provide the ultimate sensitivity of exceptional points~\cite{braunstein1996generalized}. While QFI with exceptional points has been calculated in recent works~\cite{chen2019sensitivity,zhang2019quantum}, significant disagreement still exists regarding the sensing benefit of exceptional points. 

In this Letter, we aim to clarify the contribution of exceptional points for photonic sensing. To estimate the ultimate sensitivity, we calculate QFI in different configurations of non-Hermitian photonic systems under the linear approximation. We find that QFI shows divergent behavior when at least one photonic mode is at the lasing threshold. On the other hand, QFI is bounded to finite values at exceptional points. This confirms that the fundamental reason for improved sensitivity is the lasing threshold, instead of exceptional points. However, there is one special case such that exceptional points and lasing threshold are achieved at the same time. This condition can only be achieved if the total gain and loss are balanced, thus the overall system is parity-time (PT) symmetric~\cite{el2018non,ozdemir2019parity,miri2019exceptional}. In this case, QFI diverges with a faster speed than only the single-mode lasing threshold. The divergence speed is proportional to the order of exceptional points. Therefore, we conclude that exceptional points can benefit the sensitivity, but only in combination with the lasing threshold in PT-symmetric systems. Our analysis reveals the importance of the balance between the total gain and loss to achieve enhanced sensitivity at exceptional-points, providing guidance for further experimental verification.

\begin{figure}[tb]
\centering
\includegraphics[width=0.8\columnwidth]{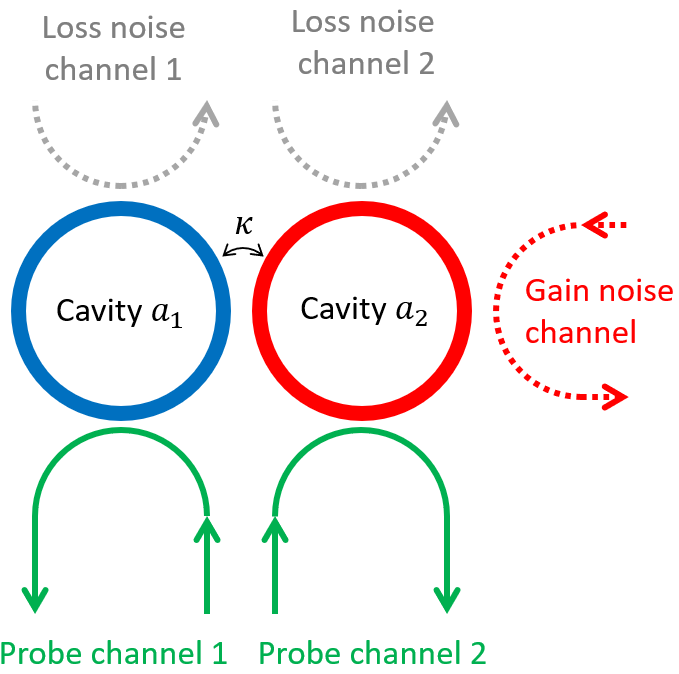}
\caption{Schematic of the photonic dimer with two coupled resonators. Probe channels (solid green) can be accessed experimentally. Intrinsic loss channels (dashed gray) and gain channel (dashed red) cannot be accessed experimentally.} 
\label{fig:scheme} 
\end{figure}

\begin{figure}[tb]
\centering
\includegraphics[width=0.8\columnwidth]{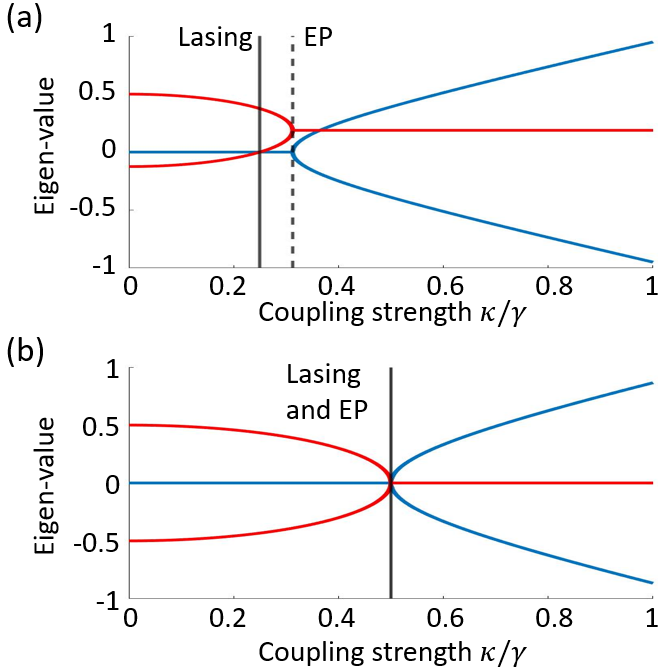}
\caption{\label{fig:1} Eigen-value real part (blue) and imaginary part (red) with (a) $g=1.25\gamma$ and (b) $g=2\gamma$. The conditions for the lasing threshold and exceptional point (EP) are labeled with solid and dashed black lines respectively. The two conditions coincide with $g=2\gamma$.} 
\label{fig:eigen}
\end{figure}

Photonic dimer, consisting of two coupled photonic cavities, is a well-established model to study exceptional points and non-Hermitian dynamics (Fig.~\ref{fig:scheme}). The classical equation of motion for this photonic system can be written as
\begin{equation}
\begin{aligned}
    &\frac{da_1}{dt}=(-i\omega_0-\frac{\gamma}{2}) a_1+i\kappa a_2 \\
    &\frac{da_2}{dt}=(-i\omega_0-\frac{\gamma-g}{2}) a_2 +i\kappa a_1 \\
\end{aligned}
\end{equation}
where $a_1$ and $a_2$ represent optical fields in each cavity, $\omega_0$ is the resonant frequency, $\gamma$ is the cavity loss, and $\kappa$ is the coupling rate between the two cavities. Here, we assume that the two cavities have the same resonant frequency and cavity loss, and optical gain $g$ is applied to the second cavity. The eigen-values are given by 
\begin{equation}
\begin{aligned}
\lambda_{\pm}=\omega_0+i\frac{g-2\gamma}{4}\pm\sqrt{\kappa^2-(\frac{g}{4})^2}
\label{Eq:eigen}
\end{aligned}
\end{equation}
With $g=4\kappa$, the two eigen-values become degenerate, corresponding to the exceptional point (Fig.~\ref{fig:eigen}a). Lasing threshold requires the imaginary part of one eigen-values to be zero. This can be obtained under two conditions. The first one is $g = \gamma+4\frac{\kappa^2}{\gamma}$, corresponding to single-mode lasing threshold without frequency detune at $\omega_0$. The second one is $g=2\gamma$ and $\kappa>\gamma/2$, corresponding to single-mode lasing threshold with frequency detune at $\omega_0\pm\sqrt{\kappa^2-(g/4)^2}$. The conditions for the exceptional point and lasing threshold can be simultaneously satisfied when $g=2\gamma=4\kappa$. This corresponds to exceptional points in PT-symmetric systems with balanced total gain and loss (Fig.~\ref{fig:eigen}b).

From quantum perspective, gain and loss are realized through phase-insensitive coupling with environments, which inevitably introduces excess noises~\cite{gardiner2004quantum}. As optical fields are amplified or attenuated with the same ratio regardless of the optical phase, additional noise channels are essential to maintain the quantum mechanical commutation relation (Fig. 1). To investigate the ultimate sensitivity limit, we use quantum Langevin equations to describe the dynamics of the coupled photonic cavities.
\begin{equation}
\begin{aligned}
    \frac{d\hat{a}_1}{dt}=(-i\omega_0-& \frac{\gamma}{2}) \hat{a}_1+i\kappa \hat{a}_2 \\
    & +\sqrt{\gamma_\mathrm{c}}\hat{a}_{\mathrm{1}}^\mathrm{in}+\sqrt{\gamma_\mathrm{i}}\hat{N}_{1}\\
    \frac{d\hat{a}_2}{dt}=(-i\omega_0-& \frac{\gamma-g}{2}) \hat{a}_2 +i\kappa \hat{a}_1 \\
    & +\sqrt{\gamma_\mathrm{c}}\hat{a}_{\mathrm{2}}^\mathrm{in}+\sqrt{\gamma_\mathrm{i}}\hat{N}_{2}-\sqrt{g}\hat{N}_g^\dagger\\
\end{aligned}
\label{Eq:QLE}
\end{equation}
where $\hat{N}_{1/2}$ is the noise operator due to the intrinsic loss, $\hat{a}_\mathrm{1/2}^\mathrm{in}$ is the input state operator from the probe channel, and the gain $g$ is realized through the gain channel with the creation operator $\hat{N}_g^\dagger$. Here we assume the two cavities have the same passive intrinsic loss ($\gamma_\mathrm{i}$) and coupling strength ($\gamma_\mathrm{c}$) with the probe channels. After performing the Fourier transform $\hat{a}(\omega)=\int\hat{a}(t)e^{-i(\omega-\omega_0)t}dt$ and converting into quadrature basis $x=\hat{a}+\hat{a}^\dagger$ and $y=-i(\hat{a}-\hat{a}^\dagger$), Eq.~(\ref{Eq:QLE}) can be rewritten as
\begin{equation}
\begin{aligned}
\begin{bmatrix}
x_1\\
x_2\\
y_1\\
y_2
\end{bmatrix}
=
G
\left(\sqrt{\gamma_\mathrm{c}}
\begin{bmatrix}
x_\mathrm{1}^\mathrm{in}\\
x_\mathrm{2}^\mathrm{in}\\
y_\mathrm{1}^\mathrm{in}\\
y_\mathrm{2}^\mathrm{in}
\end{bmatrix}
+\sqrt{\gamma_\mathrm{i}}
\begin{bmatrix}
q_{1}\\
q_{2}\\
p_{1}\\
p_{2}
\end{bmatrix}
+\sqrt{g}
\begin{bmatrix}
0\\
-p_{g}\\
0\\
q_{g}
\end{bmatrix}
\right)
\end{aligned}
\end{equation}
where ($q_{1/2}$, $p_{1/2}$) and ($q_{g}$, $p_{g}$) are the quadratures of intrinsic noise channels and the gain channel respectively, and ($x_\mathrm{1/2}^\mathrm{in}$, $y_\mathrm{1/2}^\mathrm{in}$) are the input state quadratures. The transfer function $G$ is dependent on the cavity detune $\theta=\omega-\omega_0$ 
\begin{equation}
\begin{aligned}
G^{-1}(\theta)=
\begin{bmatrix}
\frac{\gamma}{2} & 0 & \theta & \kappa \\
0 & \frac{\gamma-g}{2}&\kappa&\theta\\
-\theta&-\kappa&\frac{\gamma}{2}&0\\
-\kappa&-\theta&0&\frac{\gamma-g}{2}
\end{bmatrix}
\label{Eq:G}
\end{aligned}
\end{equation}
Then we can obtain the output state quadratures
\begin{equation}
\begin{aligned}
    \begin{bmatrix}
    x_\mathrm{1}^\mathrm{out}\\
    x_\mathrm{2}^\mathrm{out}\\
    y_\mathrm{1}^\mathrm{out}\\
    y_\mathrm{2}^\mathrm{out}
    \end{bmatrix}
    &=\sqrt{\gamma_c}
    \begin{bmatrix}
    x_1\\
    x_2\\
    y_1\\
    y_2
    \end{bmatrix}
    -
    \begin{bmatrix}
    x_\mathrm{1}^\mathrm{in}\\
    x_\mathrm{2}^\mathrm{in}\\
    y_\mathrm{1}^\mathrm{in}\\
    y_\mathrm{2}^\mathrm{in}
    \end{bmatrix}\\
    &=(\gamma_\mathrm{c} G(\theta)-I)
    \begin{bmatrix}
    x_1^\mathrm{in}\\
    x_2^\mathrm{in}\\
    y_1^\mathrm{in}\\
    y_2^\mathrm{in}
    \end{bmatrix}\\
    &+\sqrt{\gamma_c\gamma_i} G(\theta)
    \begin{bmatrix}
    q_1\\
    q_2\\
    p_1\\
    p_2
    \end{bmatrix}
    +\sqrt{\gamma_c g} G(\theta)
    \begin{bmatrix}
    0\\
    -p_g\\
    0\\
    q_g
    \end{bmatrix}
\end{aligned}
\end{equation}
where $I$ is the 4-by-4 indentify matrix. Noise quadratures ($q_{1/2}$, $p_{1/2}$) and ($q_{g}$, $p_{g}$) are averaged to zero. Therefore we can obtain the signal output.
\begin{equation}
\begin{aligned}
\mu_\mathrm{out}(\theta)\stackrel{\text{def}}{=}
    \begin{bmatrix}
    \bar{x}_\mathrm{1}^\mathrm{out}\\
    \bar{x}_\mathrm{2}^\mathrm{out}\\
    \bar{y}_\mathrm{1}^\mathrm{out}\\
    \bar{y}_\mathrm{2}^\mathrm{out}
    \end{bmatrix}
    &=(\gamma_\mathrm{c} G(\theta)-I)
    \begin{bmatrix}
    \bar{x}_\mathrm{1}^\mathrm{in}\\
    \bar{x}_\mathrm{2}^\mathrm{in}\\
    \bar{y}_\mathrm{1}^\mathrm{in}\\
    \bar{y}_\mathrm{2}^\mathrm{in}
    \end{bmatrix}\\
    &\stackrel{\text{def}}{=}(\gamma_\mathrm{c} G(\theta)-I)\;\mu_\mathrm{in}
\label{Eq:sig}
\end{aligned}
\end{equation}
The covariance matrix of the output state, which characterize the noise, can be written as
\begin{equation}
\begin{aligned}
    V_{\mathrm{out}}(\theta)&=\left(\gamma_{\mathrm{c}}G-I\right)V_{\mathrm{in}}\left(\gamma_{c}G-I\right)^T\\
    &+\gamma_{\mathrm{c}}\gamma_{\mathrm{n}}GV_{\mathrm{i}}G^{T}+\gamma_{\mathrm{c}}gGDV_{\mathrm{g}}D^{T}G^{T}
\label{Eq:cov}    
\end{aligned}
\end{equation}
where $V_{\mathrm{in}}$, $V_{\mathrm{i}}$ and $V_g$ are the covariance matrix for the input state, noise input due to intrinsic loss and gain channel, respectively. For simplicity, we assume all noise and input channels have quantum-limited vacuum noise. For optics, this corresponds to coherent state input and perfect amplification with no thermal noise. Therefore, we have
\begin{equation}
\begin{aligned}
    V_{\mathrm{in}}=V_{\mathrm{i}}=V_{\mathrm{g}}=
    \begin{bmatrix}
1 & 0 & i & 0 \\
0 & 1 & 0 & i \\
-i & 0 & 1 & 0 \\
0 & -i & 0 & 1
\end{bmatrix}
    ,\;D=\mathrm{diag}[0,-1,0,1]
\label{Eq:cov_in}
\end{aligned}
\end{equation}
As the overall photonic process is Gaussian, QFI can be directly calculated using the average signal output (Eq.~(\ref{Eq:sig})) and covariance matrix (Eq.~(\ref{Eq:cov}))~\cite{monras2013phase,jiang2014quantum}. If we use cavity detune $\theta$ as the perturbation and assume strong input, QFI can be written as
\begin{equation}
\begin{aligned}
    I(\theta)=\left(\frac{d\mu_\mathrm{out}(\theta)}{d\theta}\right)^{T}V^{-1}_\mathrm{out}(\theta)\left(\frac{d\mu_\mathrm{out}(\theta)}{d\theta}\right)
\label{Eq:QFI}
\end{aligned}
\end{equation}  
Then the sensitivity lower-bound $\delta\theta$ is determined by quantum Cramer Rao bound~\cite{braunstein1994statistical}
\begin{equation}
\begin{aligned}
    \delta\theta=1/\sqrt{I(\theta)}
\end{aligned}
\end{equation}

\begin{figure}[tb]
\centering
\includegraphics[width=0.8\columnwidth]{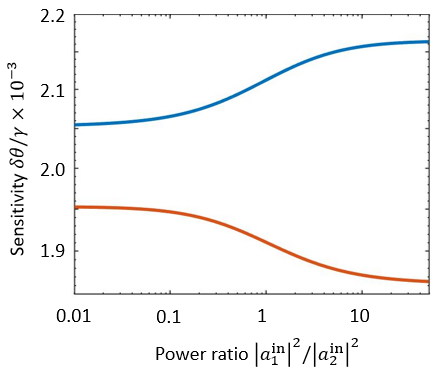}
\caption{Sensitivity dependence on input power distribution at the lasing threshold $g=1.9\gamma$ (blue) and $g=2.1\gamma$ with $\kappa$ calculated by $g=\gamma+4\kappa^2/\gamma$. Perturbation amplitude is set as $\theta=10^{-3}$.} 
\label{fig:power} 
\end{figure}

With Eq.~(\ref{Eq:sig}-\ref{Eq:cov_in}), we can see that QFI is only determined by the input probe amplitude $\mu_\mathrm{in}$ and the transfer function $G(\theta)$.
\begin{equation}
\begin{aligned}
    &I(\theta)=\\
    &\gamma_\mathrm{c}^2\left(\frac{dG}{d\theta}\mu_\mathrm{in}\right)^T
    \begin{bmatrix}
    &(\gamma_{\mathrm{c}}G-I)V_{\mathrm{in}}(\gamma_{c}G-I)^T\\
    &+\gamma_{\mathrm{c}}\gamma_{\mathrm{i}}GV_{\mathrm{in}}G^{T}\\
    &+\gamma_{\mathrm{c}}gGDV_{\mathrm{in}}D^{T}G^{T}
    \end{bmatrix}^{-1}
    \left(\frac{dG}{d\theta}\mu_\mathrm{in}\right)
\end{aligned}
\end{equation}

We first analyze the impact of the input state $\mu_\mathrm{in}$. As both the optical gain and loss are phase-insensitive, we focus on the power distribution between the two input waveguides. We fix the total input power $|\hat{a}_\mathrm{1}^\mathrm{in}|^2+|\hat{a}_\mathrm{2}^\mathrm{in}|^2$, and vary the power ratio $|\hat{a}_\mathrm{1}^\mathrm{in}|^2/|\hat{a}_\mathrm{2}^\mathrm{in}|^2$. As shown in Fig.~\ref{fig:power}, the proper choice of power ratio can improve the sensitivity. At the lasing threshold $g = \gamma+4\frac{\kappa^2}{\gamma}$, we should allocate all power to the cavity at the lasing threshold. This corresponds to all power to the active cavity $\hat{a}_2$ if $g<2\gamma$, and to the passive cavity $\hat{a}_1$ if $g>2\gamma$. For exceptional points with balanced gain and loss $g=2\gamma=4\kappa$, the sensitivity limit has no dependence on the power distribution. This is because both modes are at the lasing threshold, and both modes have equal distribution in the two cavities. Regardless of photonic parameters, the sensitivity is a smooth function without divergence. The choice of the input state does not change the sensitivity radically. Therefore, we use the balanced input $|\hat{a}_\mathrm{1}^\mathrm{in}|^2/|\hat{a}_\mathrm{2}^\mathrm{in}|^2=1$ to analyze the impact of the transfer function $G(\theta)$ on the sensitivity limit.

\begin{figure}[tb]
\centering
\includegraphics[width=0.8\columnwidth]{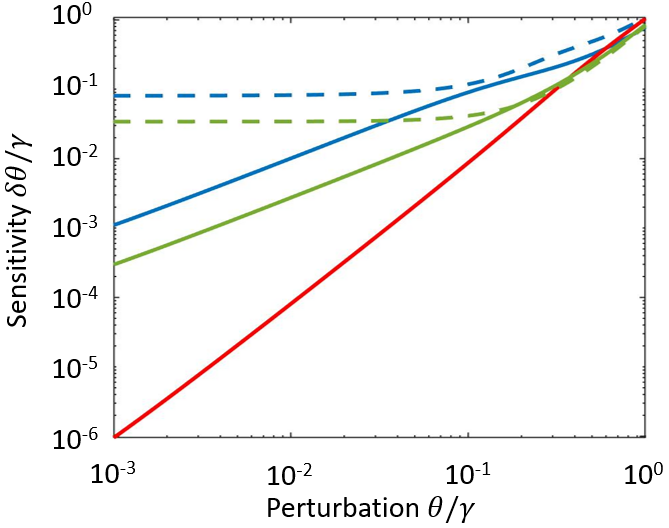}
\caption{Sensitivity dependence on perturbation amplitude. Blue dashed line: generalized exceptional point with $\kappa=g/4$ and $g=\gamma$. Green dashed line: generalized exceptional point with $\kappa=g/4$ and $g=3\gamma$. Blue solid lines: single-mode lasing threshold with $g=\gamma+4\kappa^2/\gamma$ and $g=\gamma$. Green solid lines: single-mode lasing threshold with $g=\gamma+4\kappa^2/\gamma$ and $g=3\gamma$.
Solid red line: exceptional point with balanced gain and loss $g=2\gamma=4\kappa$.} 
\label{fig:lasing1} 
\end{figure}

Different from the smooth dependence on the input state, the sensitivity can show divergent dependence on the transfer function $G(\theta)$ due to the inverse operation (Eq.~(\ref{Eq:G})). This occurs when the inverse of the transfer function has zero determinant
\begin{equation}
\begin{aligned}
\mathrm{det}[G^{-1}(\theta_0)]=
\begin{vmatrix}
\frac{\gamma}{2} & 0 & \theta_0 & \kappa \\
0 & \frac{\gamma-g}{2} & \kappa & \theta_0\\
-\theta_0 & -\kappa & \frac{\gamma}{2} & 0\\
-\kappa & -\theta_0 & 0 & \frac{\gamma-g}{2}
\end{vmatrix}
=0
\label{Eq:det}
\end{aligned}
\end{equation}
If we assume the perturbation is dispersive, $\theta_0$ can only have real values. Then we can obtain two conditions for Eq.~(\ref{Eq:det}): (i) $g = \gamma+4\frac{\kappa^2}{\gamma}\ge\gamma$ with $\theta_0=0$ and (ii) $g=2\gamma$ and $\kappa>\gamma/2$ with $\theta_0=\pm\sqrt{\kappa^2-(g/4)^2}$, corresponding to the undetuned and detuned lasing threshold identified from Eq.~(\ref{Eq:eigen}) respectively. Near these conditions, Laurent expansion around the defective point $\theta_0$ can be used to describe the dynamics of the transfer function.
\begin{equation}
\begin{aligned}
    G(\theta)=\sum_{k=0}^m C_k \cdot \frac{1}{(\theta-\theta_0)^k} \sim  C_m \cdot \frac{1}{(\theta-\theta_0)^m}
\end{aligned}
\end{equation}
where $C_k$ is the expansion constant matrix, and $m$ is the pole order. It has been shown that the sensitivity determined by quantum Cramer Rao bound has the same scaling as the transfer function~\cite{zhang2019quantum}
\begin{equation}
\begin{aligned}
    \delta\theta \sim \frac{1}{(\theta-\theta_0)^m}
\end{aligned}
\end{equation}
Then the critical issue is to obtain the pole order $m$ under different conditions. For the first condition $g = \gamma+4\frac{\kappa^2}{\gamma}\ge\gamma$, we have $m=1$ and the expansion is conducted around $\theta_0=0$
\begin{equation}
\begin{aligned}
    G(\theta)\sim  
    \begin{bmatrix}
    0 & 2\kappa\gamma & -4\kappa^2 & 0\\
    2\kappa\gamma & 0 & 0 & \gamma^2\\
    4\kappa^2 & 0 & 0 & 2\kappa\gamma\\
    0 & -\gamma^2 & 2\kappa\gamma & 0
\end{bmatrix}
    \cdot \frac{1}{(4\kappa^2-\gamma^2)} \cdot \frac{1}{\theta}
\label{Eq:Laurent_1}    
\end{aligned}
\end{equation}
Therefore, the sensitivity diverges as $\delta\theta \sim 1/\theta$, shown as blue and green solid lines in Fig.~\ref{fig:lasing1}. In contrary, the transfer function $G(\theta)$ is non-defective at the general exceptional point condition $g=4\kappa$, corresponding to the pole order $m=0$. Therefore, the sensitivity saturates at small perturbation, shown as blue and green dashed lines in Fig.~\ref{fig:lasing1}. This shows that the fundamental reason of the sensing advantage is the lasing threshold instead of the exceptional point. However, the Laurent decomposition in Eq.~(\ref{Eq:Laurent_1}) requires the condition $\gamma\neq2\kappa$.

\begin{figure}[t]
\centering
\includegraphics[width=0.8\columnwidth]{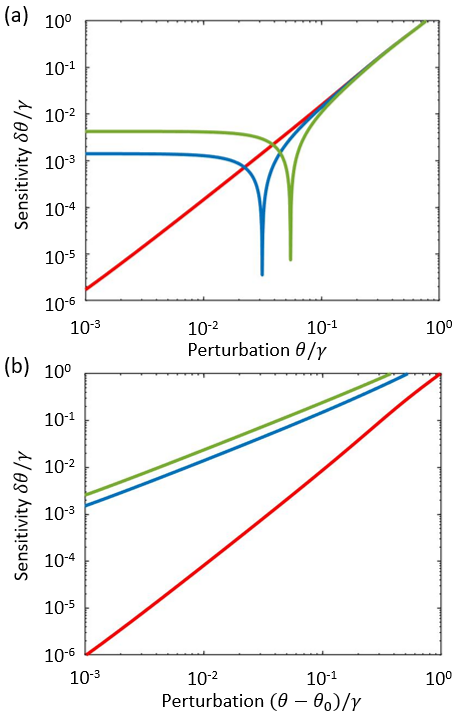}
\caption{
(a) Sensitivity dependence on original perturbation amplitude. (b) Sensitivity dependence on shifted perturbation amplitude $\theta-\sqrt{\kappa^2-(\gamma/2)^2}$. Total gain and loss are balanced $g=2\gamma$. Red: exceptional point $g=2\gamma=4\kappa$. Blue: detuned lasing threshold $\kappa=(1/2 + 10^{-3})\gamma$. Green: detuned lasing threshold $\kappa=(1/2 + 2\times10^{-3})\gamma$.
}
\label{fig:lasing2}
\end{figure}

With the condition $\gamma=2\kappa$, the lasing threshold requirement $g = \gamma+4\frac{\kappa^2}{\gamma}$ becomes $g=2\gamma=4\kappa$. Therefore, the exceptional point and lasing threshold conditions are satisfied at the same time. The photonic system is PT-symmetric with balanced total gain and loss. Then the Laurent decomposition of the transfer function gives~\cite{zhang2019quantum}
\begin{equation}
\begin{aligned}
    G(\theta)\sim 
    \begin{bmatrix}
    1 & 0 & 0 & 1\\
    0 & -1 & 1 & 0\\
    0 & -1 & 1 & 0\\
    -1 & 0 & 0 & -1
    \end{bmatrix}
    \cdot \frac{\gamma}{2} \cdot \frac{1}{\theta^2}
\end{aligned}
\end{equation}
Therefore, the pole order becomes $m=2$. The sensitivity diverges as $\delta\theta \sim 1/\theta^2$ (red solid line in Fig.~\ref{fig:lasing1}), which is faster than the undetuned single-mode lasing threshold. This shows that the exceptional point can provide sensitivity advantage on top of the lasing threshold, but only with balanced total gain and loss. On the other hand, generalized exceptional points obtained by loss gauge transformation~\cite{ozdemir2019parity} cannot exhibit sensing advantage assuming noise is limited by intrinsic processes.

With the balanced total gain and loss, the lasing threshold can also be achieved with strong mutual coupling between cavities $\kappa>\gamma/2$. Then the Laurent decomposition is conducted with frequency detune around $\theta_0=\pm\sqrt{\kappa^2-(\gamma/2)^2}$
\begin{equation}
\begin{aligned}
G(\theta)\sim
\begin{bmatrix}
-\frac{\gamma}{4} & 0 & 0 &-\frac{\kappa}{2}\\
0 & \frac{\gamma}{4} & -\frac{\kappa}{2} & 0\\
0 &\frac{\kappa}{2} & -\frac{\gamma}{4} & 0\\
\frac{\kappa}{2} & 0 & 0 & \frac{\gamma}{4}
\end{bmatrix}
\cdot \frac{1}{\sqrt{\kappa^2-\frac{\gamma^2}{4}}} \cdot \frac{1}{\theta-\theta_0}
\end{aligned}
\end{equation}
Pole order is also $m=1$. In this case, the divergent sensitivity is at non-zero frequency detune $\theta_0$ (Fig.~\ref{fig:lasing2}a). If we shift the origin point of the perturbation from 0 to $\theta_0$, we can clearly see the first-order divergence of the sensitivity $\delta\theta \sim 1/(\theta-\theta_0)$ (Fig.~\ref{fig:lasing2}b). As the optimum frequency detune $\theta_0$ can be controlled with the mutual coupling $\kappa$, it can enable novel dynamic measurement schemes that track the perturbation change with high sensitivity.

\begin{figure}[t]
\centering
\includegraphics[width=0.8\columnwidth]{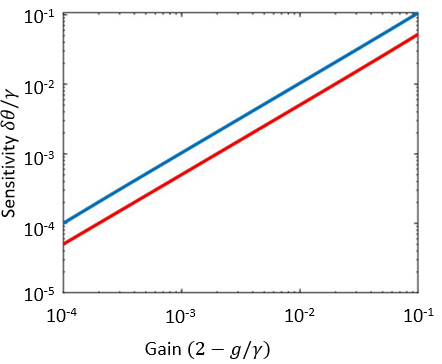}
\caption{Sensitivity dependence on the optical gain. Red: approaching the exceptional point with balanced gain and loss with $g\rightarrow 2\gamma$ and $\kappa=\gamma/2$, . Blue: approaching the detuned lasing threshold when $g\rightarrow 2\gamma$ and $\kappa=\gamma$. Perturbation amplitude is
set as $\delta\theta=10^{-5}$ away from the critical value $\theta_0=0$ and $\theta_0=\sqrt{\kappa^2-(\gamma/2)^2}$ for the red and blue lines respectively.
}
\label{fig:gain1} 
\end{figure}

Gain saturation is not considered in this linearized model. With gain saturation, we expect that the sensitivity will not show divergence as this requires infinite signal amplification and power. Practically, gain saturation can be avoided by setting the gain $g$ slightly below the divergence threshold. To evaluate the influence of such sub-threshold condition, we compare the sensitivity with varying gain $g$ in two cases (i) $\kappa=\gamma/2$ near $\theta_0=0$ and (ii) $\kappa=\gamma$ near $\theta_0=\sqrt{\kappa-(\gamma/2)^2}=\sqrt{3}\gamma/2$. At $g=2\gamma$, the two cases corresponds to the exceptional point with balanced gain and loss, and the lasing threshold with frequency detune, respectively. As shown in Fig.~\ref{fig:gain1}, the sensitivity advantage is maintained with sub-threshold gain values. This can be verified by conducting the Laurent decomposition with respect to the gain $g$. For both lasing threshold and exceptional point with balanced gain and loss, the same polar number of $m=1$ is obtained
\begin{equation}
\begin{aligned}
    G(g) \sim  \frac{1}{(g-g_0)}
\end{aligned}
\end{equation}
when we set $\theta=\theta_0$. The corresponding critical perturbation and gain values are (i) $\theta_0=0$ and $g_0=\gamma+4\kappa^2/\gamma$ for the undetuned lasing threshold (ii) $\theta_0=0$ and $g_0=2\gamma=4\kappa$ for the exceptional point with balanced gain and loss, and (iii) $\theta_0=\sqrt{\kappa^2-(\gamma/2)^2}$ and $g_0=2\gamma$ for the detune lasing threshold.

In conclusion, we have analyzed the contribution of the lasing threshold and exceptional point for photonic sensing. The sensitivity limit is obtained by calculating QFI in the linear model. We show that the sensitivity shows divergence at the lasing threshold, but smooth behavior at exceptional points. Therefore, the sensitivity improvement is a result of the lasing threshold instead of the exceptional point. While the exceptional point alone does not offer advantage, the optimum sensitivity is achieved if the exceptional point and lasing threshold conditions are satisfied at the same time. This corresponds to the exceptional point in PT-symmetric systems with balanced total gain and loss. Under this condition, the sensitivity shows a faster divergence speed than single-mode lasing threshold. Therefore, the balance between the total gain and loss is the key to verify the contribution of exceptional points. Besides clarifying the contribution of exceptional points, our work also provides the important guidance to experimentally realize exceptional-point enhanced photonic sensing.

\paragraph*{Acknowledgments}This work was supported by Office of Naval Research (N00014-19-1-2190), and National Science Foundation Grants No. ECCS-1842559 and No. CCF-1907918.

\bibliography{Ref.bib}

\end{document}